# Topmetal-M: a novel pixel sensor for compact tracking applications

Weiping Ren[a], Wei Zhou[b], Bihui You[a], Ni Fang[a], Yan Wang[a], Haibo Yang[b, c], Honglin Zhang[b, c], Yao Wang[b], Jun Liu[a], Xianqin Li[b], Ping Yang[a], Le Xiao[a], Yuezhao Zhang[b], Xiangru Qu[b, c], Shuguang Zou[d], Guangming Huang[a], Hua Pei[a], Fan Shen[a], Dong Wang[a], Xiaoyang Niu[b], Yuan Mei[e], Yubo Han[c, f, g], Chaosong Gao[a, *], Xiangming Sun[a, *] and Chengxin Zhao[b, c, *]

[a] *PLAC, Key Laboratory of Quark & Lepton Physics (MOE), Central China Normal University, Wuhan, 430079, China*

[b] *Institute of Modern Physics (IMP), Chinese Academy of Sciences, Lanzhou, 730000, China*

[c] *School of Nuclear Science and Technology, University of Chinese Academy of Sciences, Beijing, 100049, China*

[d] *College of Information Science and Engineering, Henan University of Technology, Zhengzhou, 450001, China*

[e] *Nuclear Science Division, Lawrence Berkeley National Laboratory, Berkeley, California, 94720, USA*

[f] *Institute of High Energy Physics (IHEP), Beijing, 100049, China*

[g] *State Key Laboratory of Particle Detection and Electronics, Beijing, 100049, China*

## Abstract

The Topmetal-M is a large area pixel sensor (18 mm × 23 mm) prototype fabricated in a new 130 nm high-resistivity CMOS process in 2019. It contains 400 rows × 512 columns square pixels with the pitch of 40 μm. In Topmetal-M, a novel charge collection method combing the Monolithic Active Pixel Sensor (MAPS) and the *Topmetal* sensor has been proposed for the first time. Both the ionized charge deposited by the particle in the sensor and along the track over the sensor can be collected. The in-pixel circuit mainly consists of a low-noise charge sensitive amplifier to establish the signal for the energy reconstruction, and a discriminator with a Time-to-Amplitude Converter (TAC) for the Time of Arrival (TOA) measurement. With this mechanism, the trajectory, particle hit position, energy and arrival time of the particle can be measured. The analog signal from each pixel is accessible through time-shared multiplexing over the entire pixel array. This paper will discuss the design and preliminary test results of the Topmetal-M sensor.

**Keywords:** CMOS pixel sensor, tracking detector, *Topmetal*, MAPS

[*]Corresponding authors.

Email addresses:

chaosonggao@mail.ccnu.edu.cn (Chaosong Gao), sphy2007@126.com (Xiangming Sun), chengxin.zhao@impcas.ac.cn (Chengxin Zhao)

# 1. Introduction

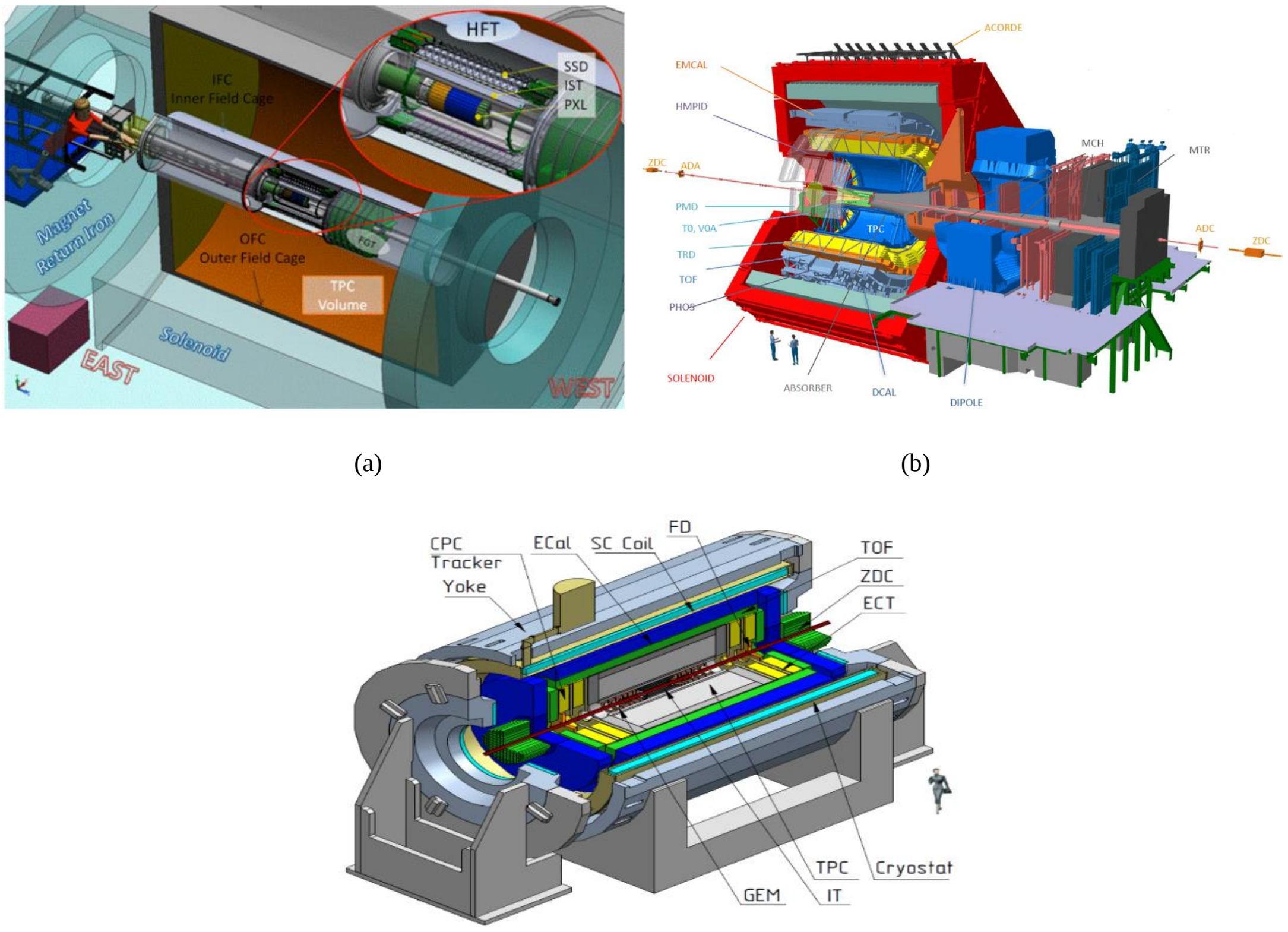

(a) (b)

(c)

Fig. 1 Tracking detectors: (a) STAR HFT detector. (b) ALICE ITS2 detector. (c) NICA MPD IT detector.

In the field of particle experiments which require the detection of charged particles, the inner tracking detectors [1][2][3] usually consists of several layers of silicon detectors for vertex construction and the TPC detector for trajectory measurement, as shown in Figure 1 (a), (b) and (c). However, many applications give only a small space for particle tracking, where small scale and compact systems are needed. The examples at the leading heavy-ion research platform in China - the Heavy Ion Research Facility at Lanzhou (HIRFL) are the beam diagnostic in the RIBLL2 [4], the real-time monitoring system at the terminal of single event effects [5], the recoil detector in the Internal Target Facility [6] and the beam monitor at the cancer therapy terminal [7][8]. Also, particle energy and timing measurement is highly demanded for accurate tracking.

Thus, a novel pixel sensor, the Topmetal-M, has been designed to provide a new option for compact tracking applications. As shown in Fig. 2 and Fig. 3, the Topmetal-M is a CMOS monolithic pixel detector, which integrates the sensing node array and the readout circuit into a single chip. Each sensing node combines the charge collecting node from the Monolithic Active Pixel Sensor (MAPS) pixel detector and the *Topmetal* sensor from the *Topmetal* pixel detectors [9][10][11]. Due to excellent spatial resolution, fast timing response and low material budget, MAPS becomes one of the most advanced technologies for particle hit measurement. The *Topmetal* pixel detectors are direct charge collectors, which have proved to be promising for TPC detectors [12]. The sensor has been thinned down to 100 μm from the backside. The charge deposited by a particle passing through the Topmetal-M will be collected by the charge collection diode. The charge generated in the ionized gas along the particle trajectory will drift in the electric field towards the Topmetal-M surface and directly collected by the exposed topmost metal layer of each pixel (the *Topmetal* sensor itself). With this mechanism, the Topmetal-M can reconstruct the trajectory, particle hit, energy and arrival time of the incident particles. This provides a feasible way to realize compact tracking detectors for small scale devices. Another important significance of the Topmetal-M is that the chip is fabricated with a new 130 nm HR (High-Resistivity) CMOS process. This aims to study the feasibility of introducing a new CMOS process to design pixel sensor for charged particles. In this paper, the design and some preliminary test results of the Topmetal-M pixel sensor will be discussed.

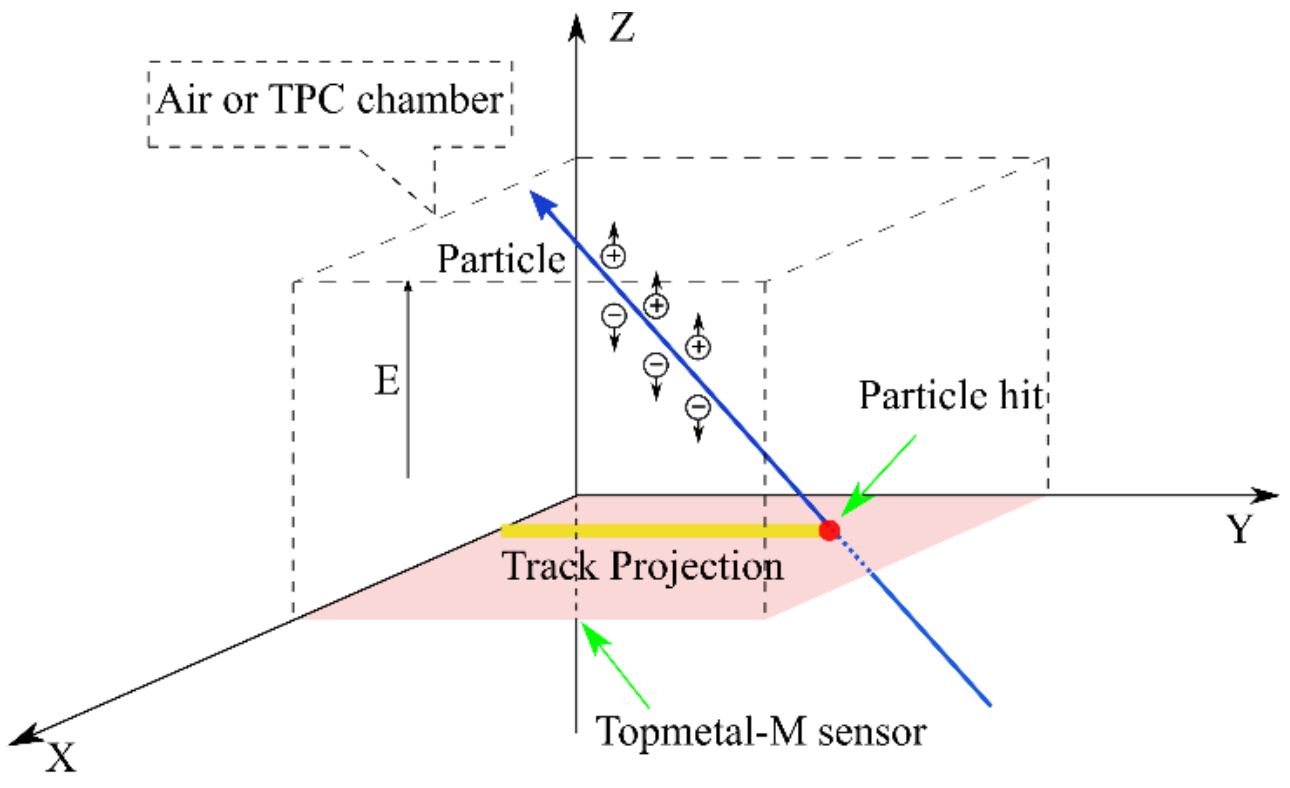

Fig. 2 Topmetal-M working principle.

## 2. Overall Architecture

Table. 1 Key parameter of the Topmetal-M sensor.

| **Parameters** | **Topmetal-M** |
|---|---|
| Process | HR (High-Resistivity) 130 nm CMOS |
| Chip Size | 18 mm × 23 mm |
| Matrix | 400 × 512 |
| Pixel Pitch | 40 × 40 μm$^2$ |
| Frame Rate | 1.5625 kHz/frame |

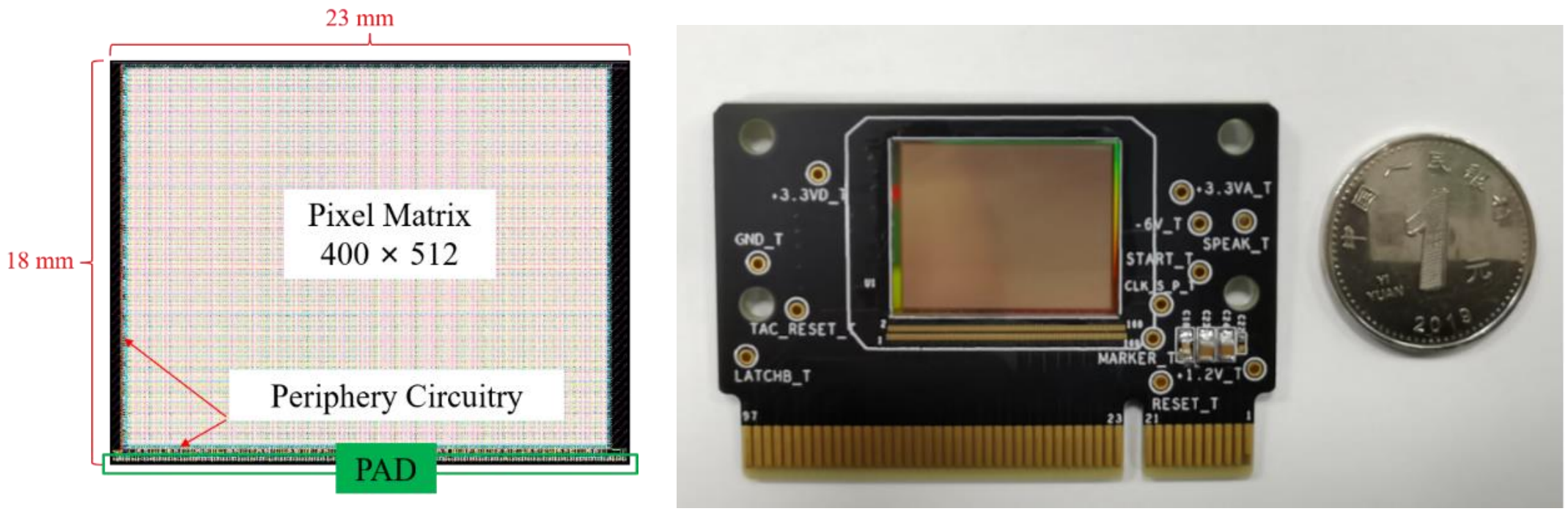

(a) (b)

Fig. 3 (a) Layout of the Topmetal-M sensor. (b) The picture of the Topmetal-M sensor.

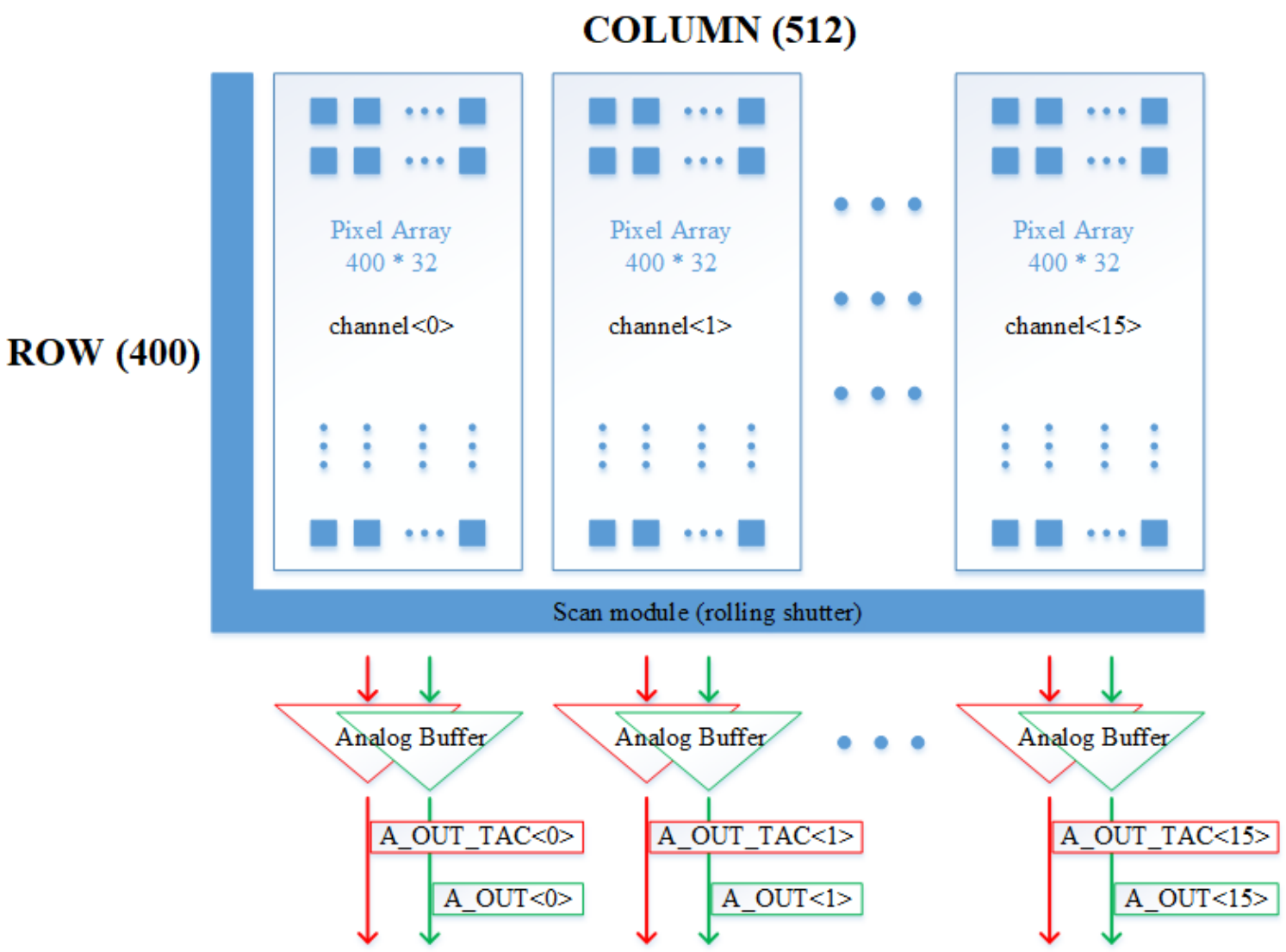

Fig. 4 Block diagram of the Topmetal-M.

The 130 nm HR CMOS (> 1 KΩ • cm) process offers a deep p-well underneath the PMOS n-well. This prevents the n-well from competing for the charge collection with the designated charge collection electrode. Hence, the full CMOS circuits can be used in the pixel design. The total size of the Topmetal-M sensor is 18 mm × 23 mm including a pixel matrix of 400 (row) × 512 (column) with the pixel pitch of 40 μm and the periphery circuit. The sensitive area is around 16 × 20.48 mm$^2$. A time-shared multiplexing readout is applied in this design and the readout frame rate can reach 1.5625 kHz/frame. A summary of the main features of Topmetal-M is given in Table. 1. The layout of the whole chip is shown in Fig. 3 (a) and the picture of the chip is shown in Fig. 3 (b). The periphery circuit is placed at the left and bottom of the pixel matrix. All the PADs are located at the bottom of the chip to make it easy to be assembled in multi-chip applications.

The overall design block diagram of the Topmetal-M is shown in Fig. 4. The pixel array is divided into 16 sub-arrays for efficient analog readout and each sub-array contains 400 × 32 pixels. The pixel unit can record the energy information carried by the incident particles, as well as the information of the arrival time. There are two dedicated analog output buffers in each sub-array. A single external clock drives the analog row or column multiplexing readout circuitry in the 16 sub-arrays synchronously. The energy and time information recorded by each pixel will be sent off the chip through the analog buffer as signal A_OUT and A_OUT_TAC. These analog signals are transmitted via coax cables to four 14-bit high-speed ADCs on the readout board. As the time of writing, a 14-bit column-parallel ADC for use inside MAPS is being developed and the first 5-bit prototype is presented in [13].

## 3. Sensor Design and Operations

The pixel structure of the Topmetal-M sensor is shown in Fig. 5. Each pixel consists of the charge collection diode, the *Topmetal*, the Charge Sensitive Amplifier (CSA), the voltage comparator, the peak holding circuit (Peak Holding), the Time to Amplitude Converter (TAC), the two-stage source follower circuit and the row selection switches. In each column, all the outputs of the row selection switches are connected to the column selection switch. In each sub-array, the output nodes of all the column switches are connected to the current source of the second stage source follower and fed into an analog buffer. The charge collected by the two sensors, the *Topmetal* and the charge collection diode, is amplified by the CSA and then split into the energy path and the timing path. The *Topmetal* is connected directly to the CSA. The charge collection diode is connected to the CSA by AC coupling. This prevents the two sensors from interfering with each other. Details on these two sensors will be introduced in section 3.1. In the energy path, the output of the CSA is fed into a source follower and the analog buffer through a time-shared multiplexing readout. The design of the energy path will be discussed in section 3.2. In the timing path, the signal goes through a comparator and then controls the start time of the TAC charging. The output of the TAC is also read out using the same way as the energy path. The design of the timing path will be discussed in section 3.3. The readout circuitry will be described in detail in section 3.4.

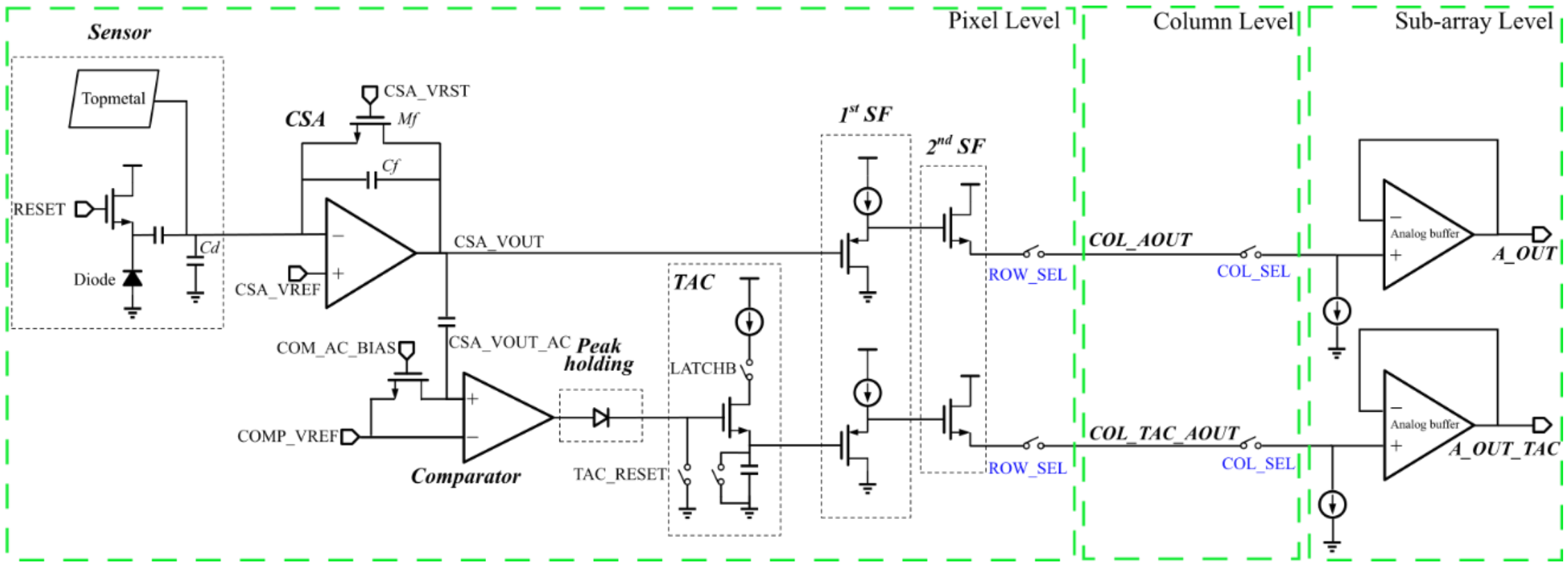

Fig. 5 The pixel structure of the Topmetal-M sensor.

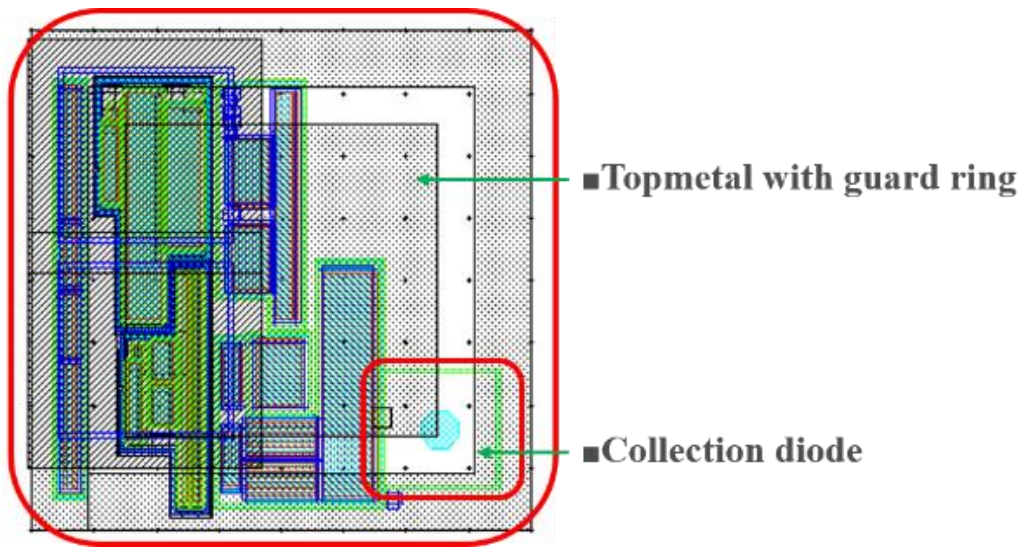

(a)

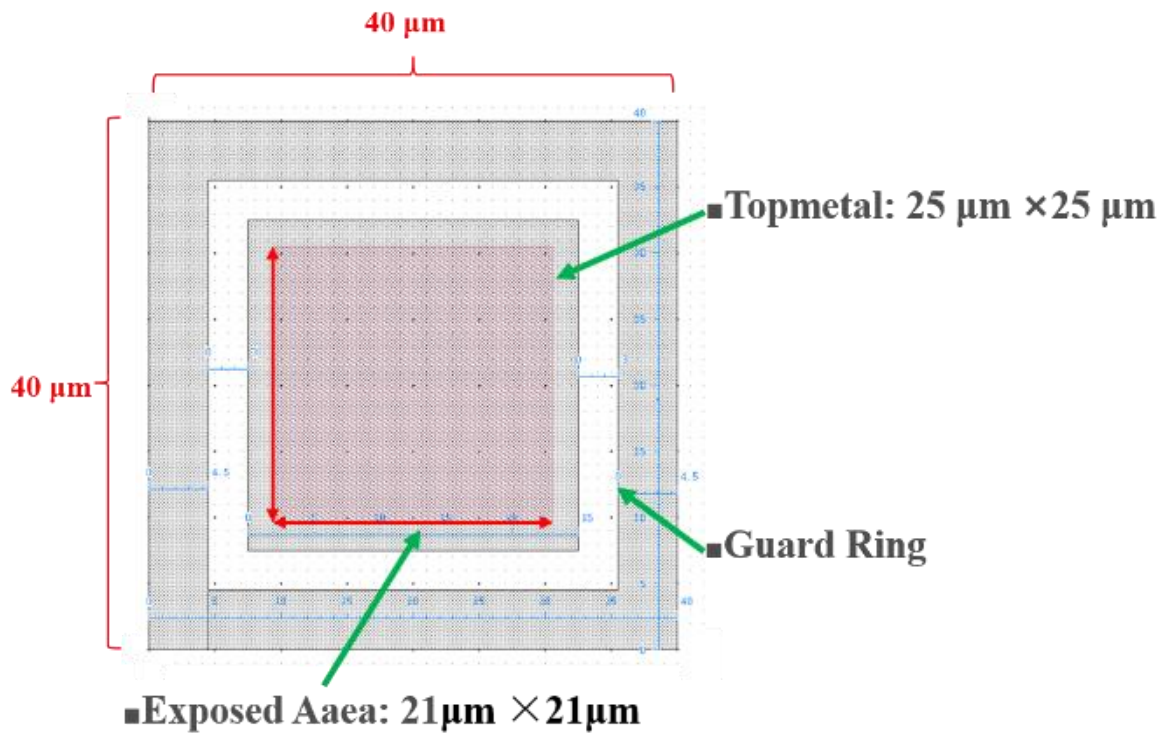

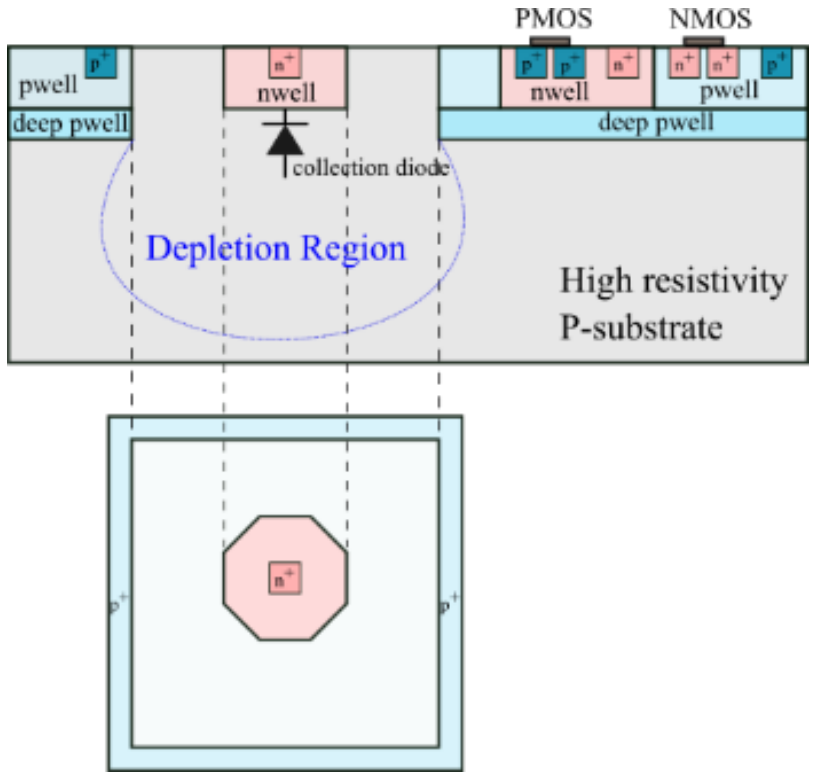

(b) (c)

Fig. 6 The sensors of Topmetal-M: (a) Location of the two kinds of sensors in one pixel. The *Topmetal* is a patch of metal of the topmost layer and it is located at the center of the pixel. The charge collection diode is at the bottom right corner of the pixel. (b) *Topmetal* with a guard ring. (c) Cross-section of MAPS and the top view of the charge collection diode structure.

### *3.1. The sensors*

The design of *Topmetal* and a pixel charge collection diode is shown in Fig. 6 (a). The *Topmetal* (Fig. 6 (b)) working principle is based on a patch of the topmost metal layer acting as a charge collection electrode, placed in each pixel cell. It is implemented as a probe-PAD located at the center of each pixel and it is directly connected to a CSA. Total size of the sensor is 25 ×25 μm$^2$ and the exposed non-insulated area is 21 ×21

μm$^2$. Each *Topmetal* is surrounded by a guard ring with the same metal layer, which is covered by an insulating layer. The space between the guard ring and the *Topmetal* is 3 μm, which is the minimum possible value under the layout design rule. All the guard rings in the pixel matrix are connected to a PAD guard-ring. With the exposed non-insulated area, the *Topmetal* collects charges directly from the surrounding media. Hence, there is no detector leakage current in the sensor. A certain voltage lower than that of *Topmetal* will be applied to the guard-ring to form a focusing electric field in each pixel cell, which will improve the charge collection efficiency. In normal working mode, the guard-ring is connected to the ground as the cathode. The *Topmetal* works at the reference voltage of the CSA, which works as the anode to collect the negative charge. The guard ring surrounding the *Topmetal* forms a coupling capacitance $C_{couple}$ of 4.7 fF. The guard ring of each pixel can be applied with an external voltage signal to emulate the electrons generated by particle hits. While testing the Topmetal-M, this voltage signal is supplied with a signal generator to characterize the performance of the CSA and the TAC.

While selecting the technology for fabricating the Topmetal-M, there were two options: the HV (High Voltage) CMOS technologies and the HR CMOS technologies. The HV CMOS technologies are attractive in view of radiation hardness since a high voltage can be applied to the sensor bulk. However, HR CMOS technologies can provide low power consumption, low threshold and fast timing since the sensor capacitance can be minimized. For these reasons, the HR CMOS technologies are more suitable for Topmetal-M. The cross-section view of the HR MAPS process and the top view of the charge collection diode in the Topmetal-M are shown in Fig. 6 (c). The charge collection is formed by the n-well (cathode) and the high resistivity P-substrate (anode) junction diode（Fig. 6 (c)). The p-n junction is reverse biased to create a depletion region around the n-well, thus it works as a collection diode to collect the negative charge. And it is possible to apply a reverse substrate bias voltage to increase the depletion region volume and reduce the collection diode junction capacitance. When a particle passes through the bulk, pairs of electron-holes are created by ionization. The electrons diffuse towards the collecting electrode. Electron-hole pairs created inside the depletion region or arriving by diffusion are collected. The spacing between n-well and surrounding p-well reduces the sidewall junction capacitance. A special feature of this technology is that the PMOS n-well is shielded from the high resistivity P-substrate by the deep p-well, which prevents it from collecting signal charge in place of the n-well collection electrode. This allows for implementing PMOS transistor in the pixel, which avoids the limitation on selecting the architecture for the in-pixel circuit.

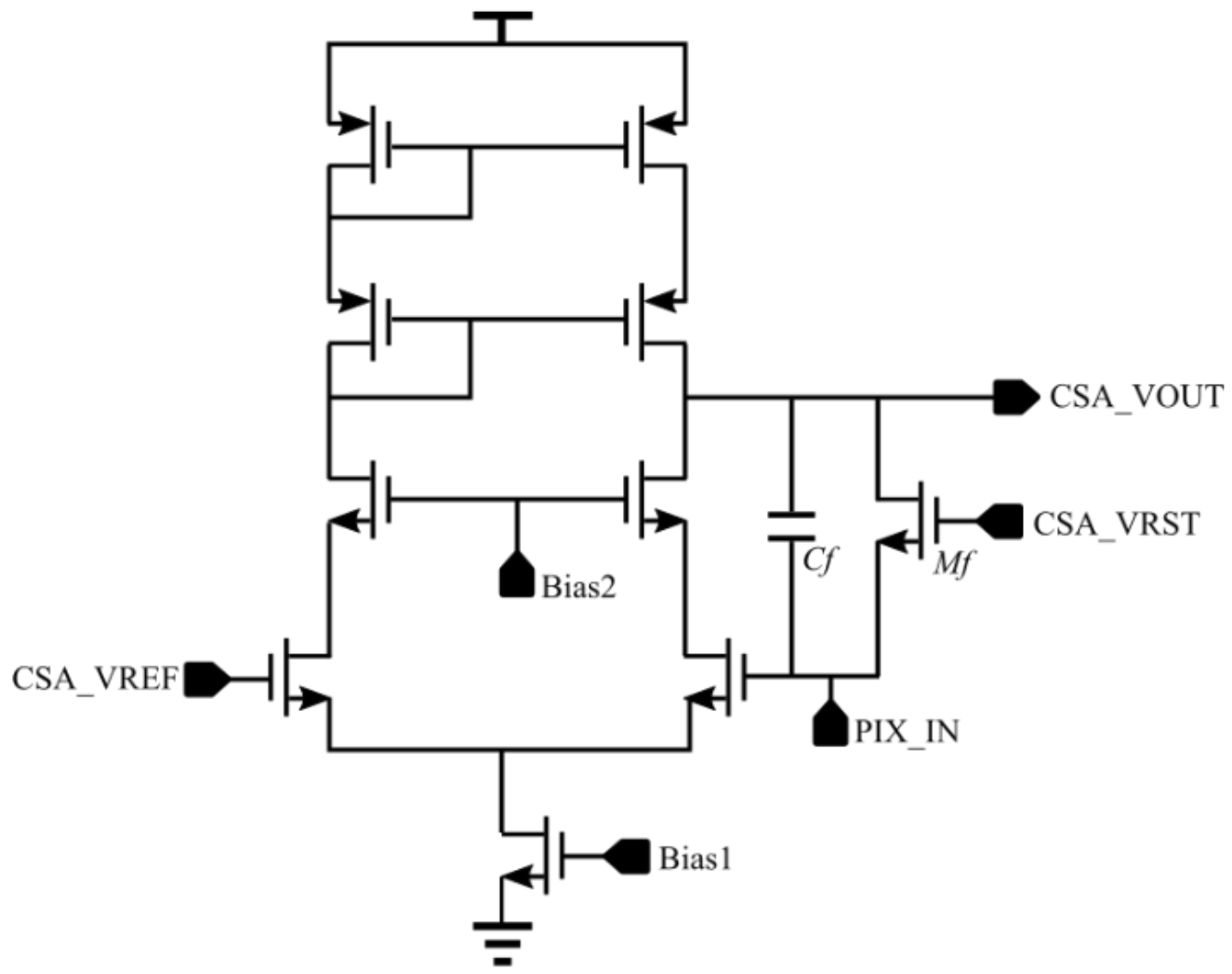

Fig. 7 The differential operation amplifier in the CSA.

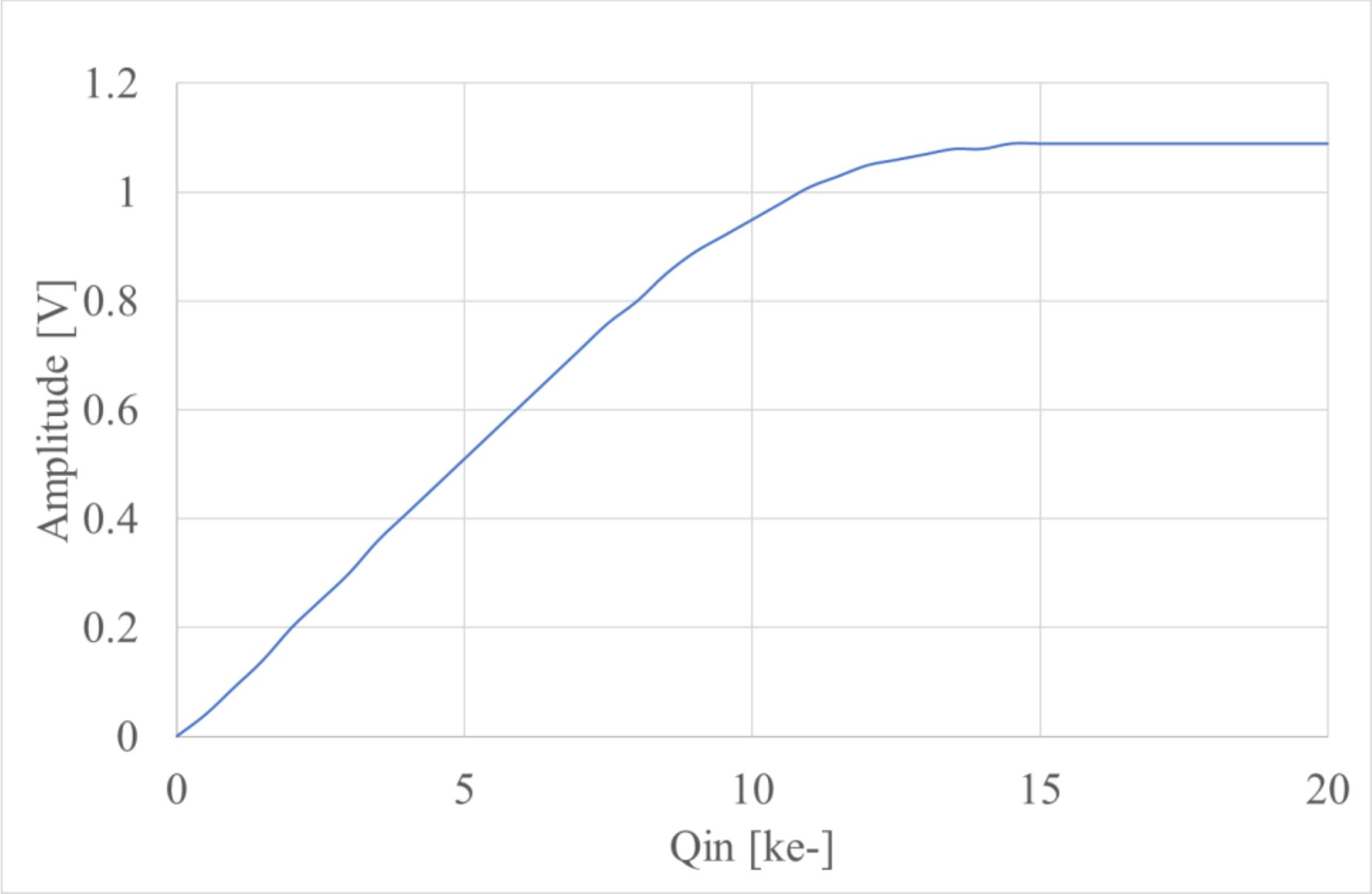

Fig. 8 The dynamic range of the charge detection of the CSA.

### *3.2. Energy measurement*

The major unit in the energy path is the CSA, which is a differential telescopic cascode amplifier architecture with a load of a cascade PMOS transistor, as shown in Fig. 7. The telescopic cascode amplifier is promising to be with low noise, which is important for our applications. The feedback capacitance $C_f$ is formed by a parasitic capacitance between two metal layers with a value of 1 fF so that the CSA conversion gain is high enough to guarantee low noise performance. The operating current of the CSA is set as 150 nA.

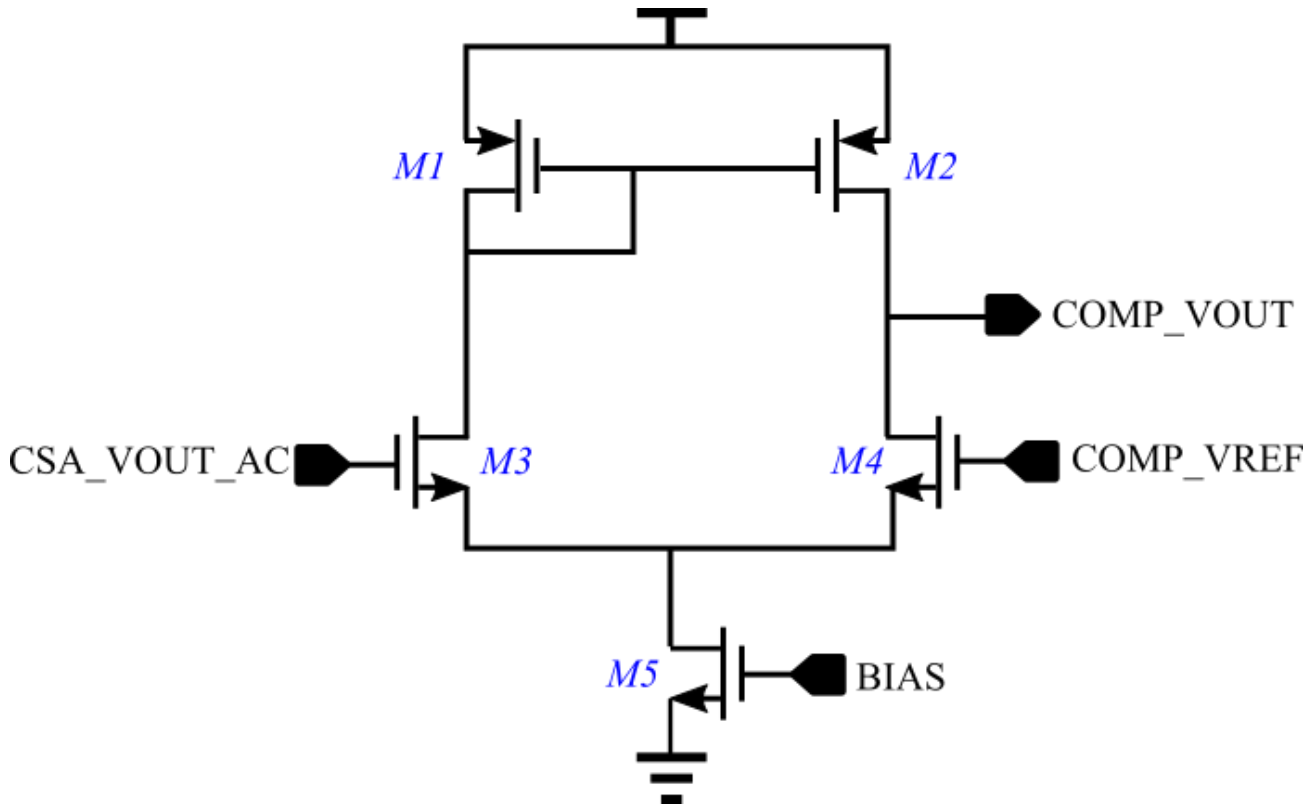

Fig. 9 The comparator structure in the pixel.

Due to its limited driving capability, the CSA is prone to be affected by the activities of the row and column selection. Hence, two stages of source follower have been used in the pixel design. The first stage of the source follower can isolate the CSA from the switching noise during readout. This reduces the dynamic range of the CSA output, but increases its stability. Fig. 8 shows the linearity of the CSA with the input charge range 0~14k $e^-$ which satisfies the requirement for this prototype version of Topmetal-M. The linearity shows good performance with $\sigma$ = 7.80 mV in the range 0 ~ 8k $e^-$. Between 8k $e^-$ and 14k $e^-$, the output amplitude still increases with the input charge while it is not linear. A calibration is applied to fully exploit the non-linear region. The charge conversion gain is about 78.6 μV/$e^-$ and the output dynamic range is 1.1 V. The decay time of the CSA could be adjusted to ensure that the peak value can be sampled during the readout of one frame. The decay time of the CSA output can be adjusted by the resistance ($M_f$, shown in Fig. 7) of the feedback transistor, which is controlled by the grid voltage of the transistor. The noise was simulated using the noise analysis tool in the spectra simulator. The simulated Equivalent Noise Charge (ENC) in this design is about 18 $e^-$.

### *3.3. The arrival time measurement*

The output of the CSA is sent to the comparator for processing (Fig. 5) by AC coupling to solve the offset problem of the CSA. A baseline restorer follows the AC coupling for signal processing. A few baseline restorer circuits [14][15][16] have been proposed in some applications. In our design, an NMOS transistor is applied between the positive and the negative input nodes of the comparator (Fig. 5). Therefore, the baseline after the AC coupling is the same as the reference voltage of the comparator. The differential pair of transistors (M1 and M2) in the comparator has been designed to be slightly unbalanced. The threshold of the comparator is formed by setting the length of the two transistors with different values. The simulation shows that the threshold is about 300 $e^-$. After the comparator, the peaking holding circuit will keep at a high level even though the comparator has already flipped from high to low. This high level will be released until the reset signal of the TAC (TAC_RESET) arrives. To minimize the size of the TAC, a light structure that consists of only 5 transistors and a capacitor has been used. This design can reduce the size of the pixel and therefore improve the spatial resolution.

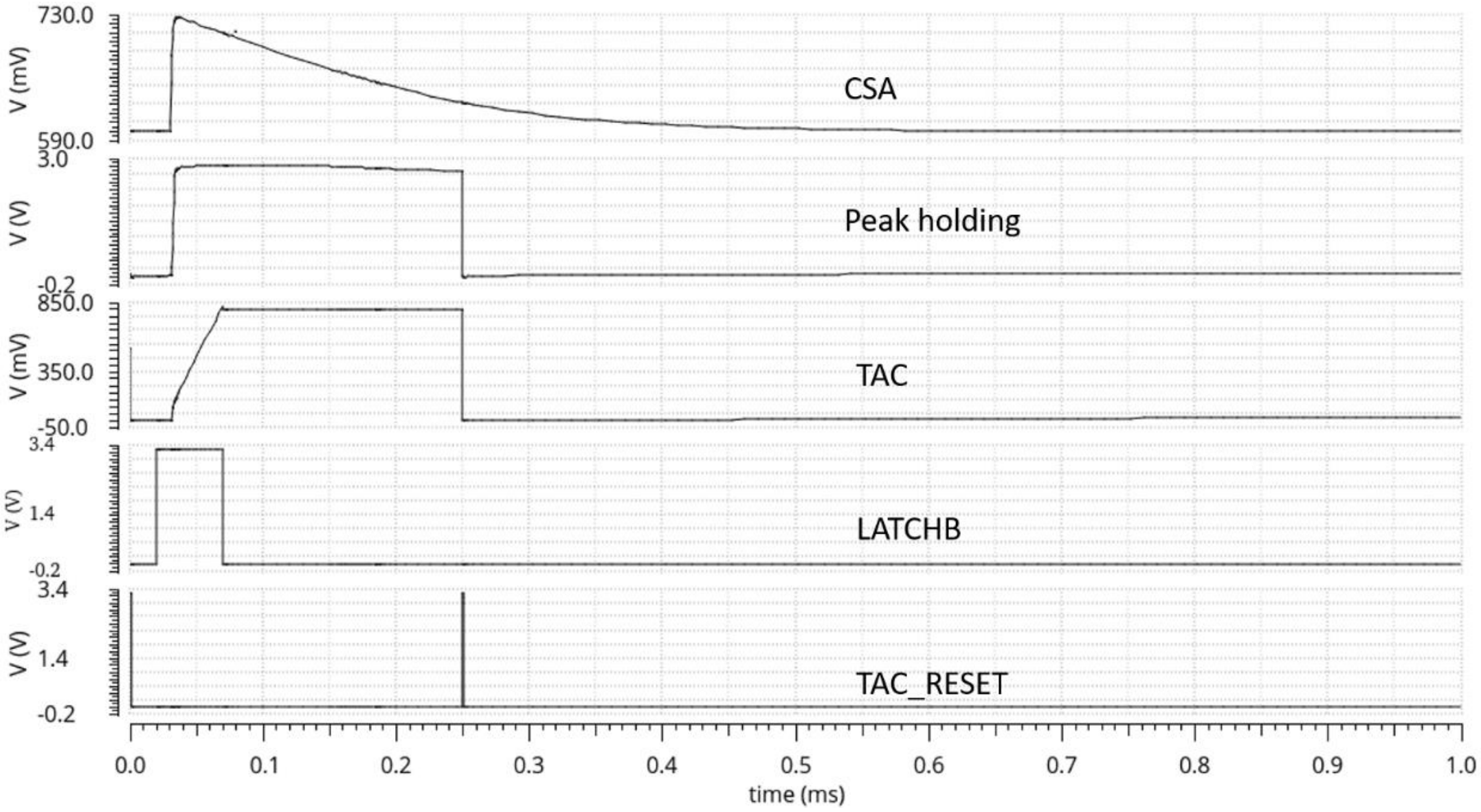

Fig. 10 The signal waveform of the pixel cell.

The procedure of timing measurement in a pixel is demonstrated in Fig. 10. Assuming enough energy is deposited into the sensor by a particle at $T_{Arrival}$, it will take a rather short time of $T_T$ to turn on the timing measurement in the TAC. During $T_T$, the output of the CSA will exceed the threshold of the comparator and the output of the comparator will become high. This output will be captured and stored by the peak holding circuit and also turn on the switch to charge the capacitance in the TAC. LATCHB is the switch of the TAC. Duration between switching on ($T_1$) and switching off ($T_2$) of the TAC is the dynamic range. The charging time ($T_{TAC}$) of the capacitance can be derived from the output voltage of the TAC ($V_{TAC}$,). The time of energy deposition ($T_{Arrival}$) can be calculated as $T_{Arrival} = T_2 - T_{TAC} - T_T$. Timing measurement can be achieved only if the $T_{Arrival}$ locates in-between $T_1$ and $T_2$. Also, the time between $T_1$ to $T_2$ should be smaller than the saturation time of charging the capacitance in the TAC. While TAC_RESET is asserted, the capacitance and the output of the peak hold circuit will be reset.

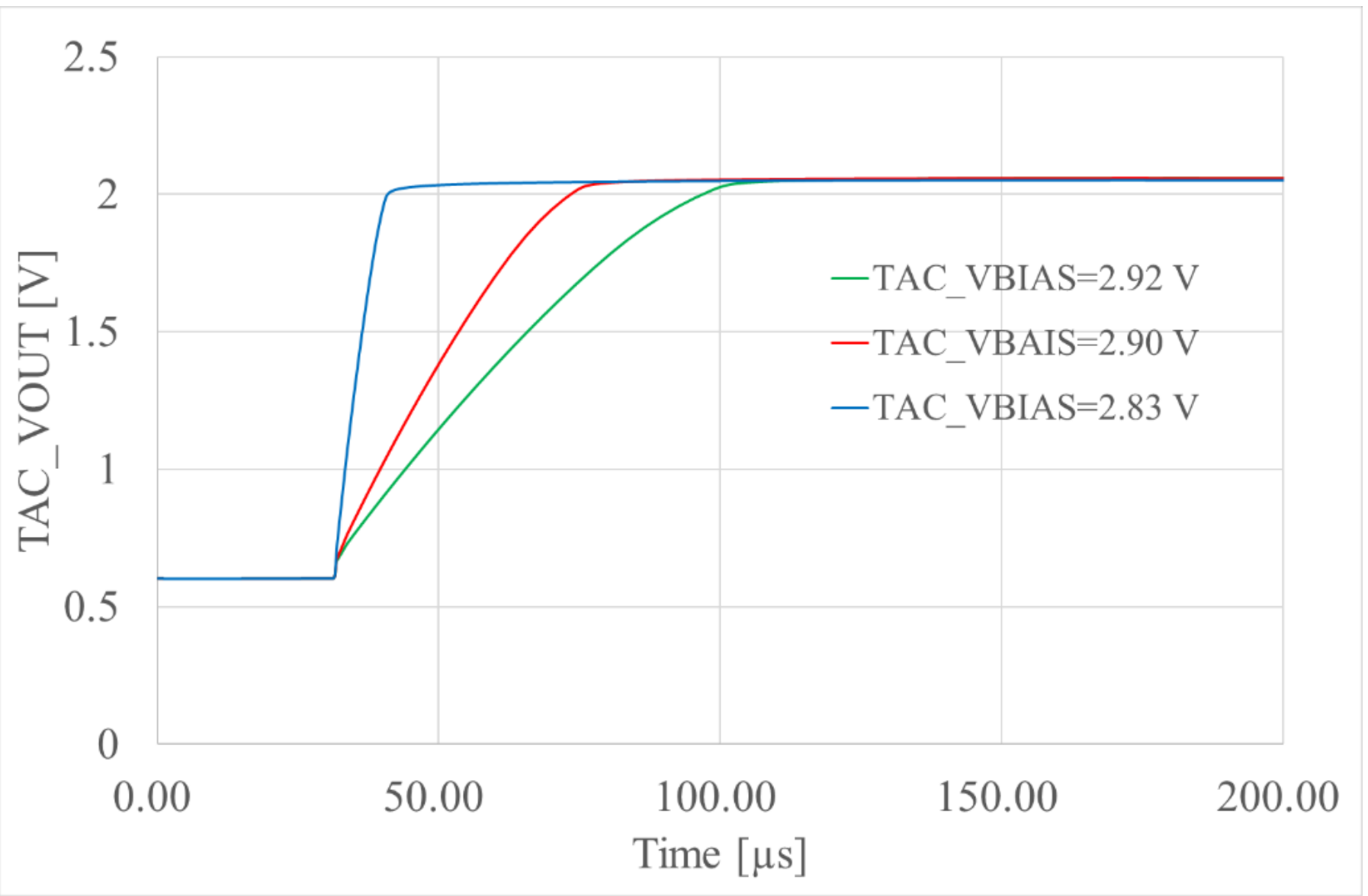

Fig. 11 The amplitude of the TAC vs the charging duration under 3 different bias voltages.

By tuning the bias voltage (TAC_VBIAS in Fig. 11) of the current source to change the charging current, the measurement range of the TAC can be adjusted for different applications. Fig. 11 shows the output of the TAC under three typical bias voltages, where the range of 10 µs, 40 µs, 75 µs and the amplitude of 1.45 V are obtained. In the readout board, a 14-bit ADC with a dynamic range of -1 V to 1 V is used to sample the output of the TAC. In this condition, the time resolution of the TAC is calculated as 0.84 ns, 3.36 ns and 6.31 ns. The reason for adopting this simple TAC architecture is to deal with the restricted constraints on the size of the pixel. The typical non-linearity of this TAC topology can be calibrated as to dedicated applications.

### *3.4. Readout*

The analog outputs from the CSA and TAC in each pixel are fed into a two-stage source follower and then transmitted out by another analog buffer. A time-shared multiplexing scheme is used to read out the entire array. As shown in Fig. 5, each analog output signal in the pixel has the 1$^{st}$ and 2$^{nd}$ source follower and the row switch (ROW_SEL). Each column shares one common switch (COL_SEL) and each of the 16 sub-array has one common output buffer. The first stage is to isolate the CSA from the switching noise in column and row selection, since the CSA has very limited driving strength and is easily to be affected. The current of the first stage source follower is set as 2 μA. The second stage source follower is only for the pixel selection in multiplexing.

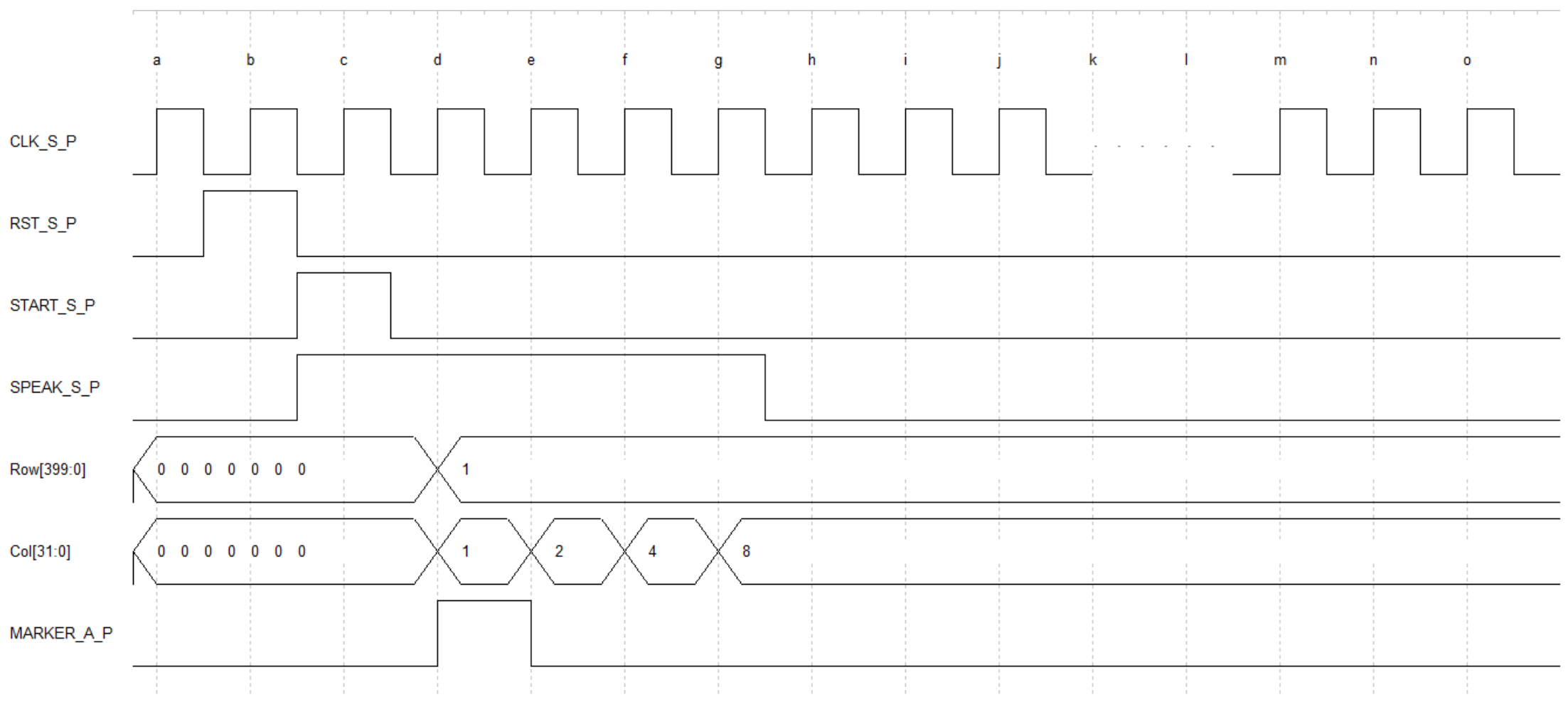

Fig. 12 The timing sequence of the Scan module in one region.

A Scan Module controls the row switch (ROW_SEL) and the column switch (COL_SEL) to select one pixel at a time for the analog readout. The timing diagram of the scan sequence in one of the 16 sub-arrays is shown in Fig. 12. The Scan Module can either scan all the pixels one by one or stop at a fixed pixel. When it recognizes the rising edge of the start signal START_S_P and the signal SPEAK_S_P is high, it begins scanning at the second rising edge of the clock. When the signal SPEAK_S_P becomes low, it will stop at the current pixel. The MAKER_A_P is the flag signal that appears when the first pixel is scanned. The first three pixels in each region are designed as marker pixels. The output of the pixels is a constant sequence as "low, high, low", which can be compared with the MAKER_A_P for synchronization.

## 4. Preliminary Test Results

### *4.1. Test setup*

Preliminary tests have been performed on the Topmetal-M. This aims to verify and study the design of this 130nm HR process. Fig. 13 shows the test system, which includes a bonding board with a Topmetal-M sensor, a readout board, an oscilloscope, a signal generator, a computer, and a power supply. The communication between the readout board and the computer is through TCP / IP protocol. The readout board is responsible for processing the control commands and reading of data from Topmetal-M. A signal generator was used to provide a voltage pulse on the guard-ring of the pixels in the Topmetal-M to characterize the performance.

All the following tests have been conducted at room temperature in ambient air and the reverse bias of the Topmetal-M is set to 0 V.

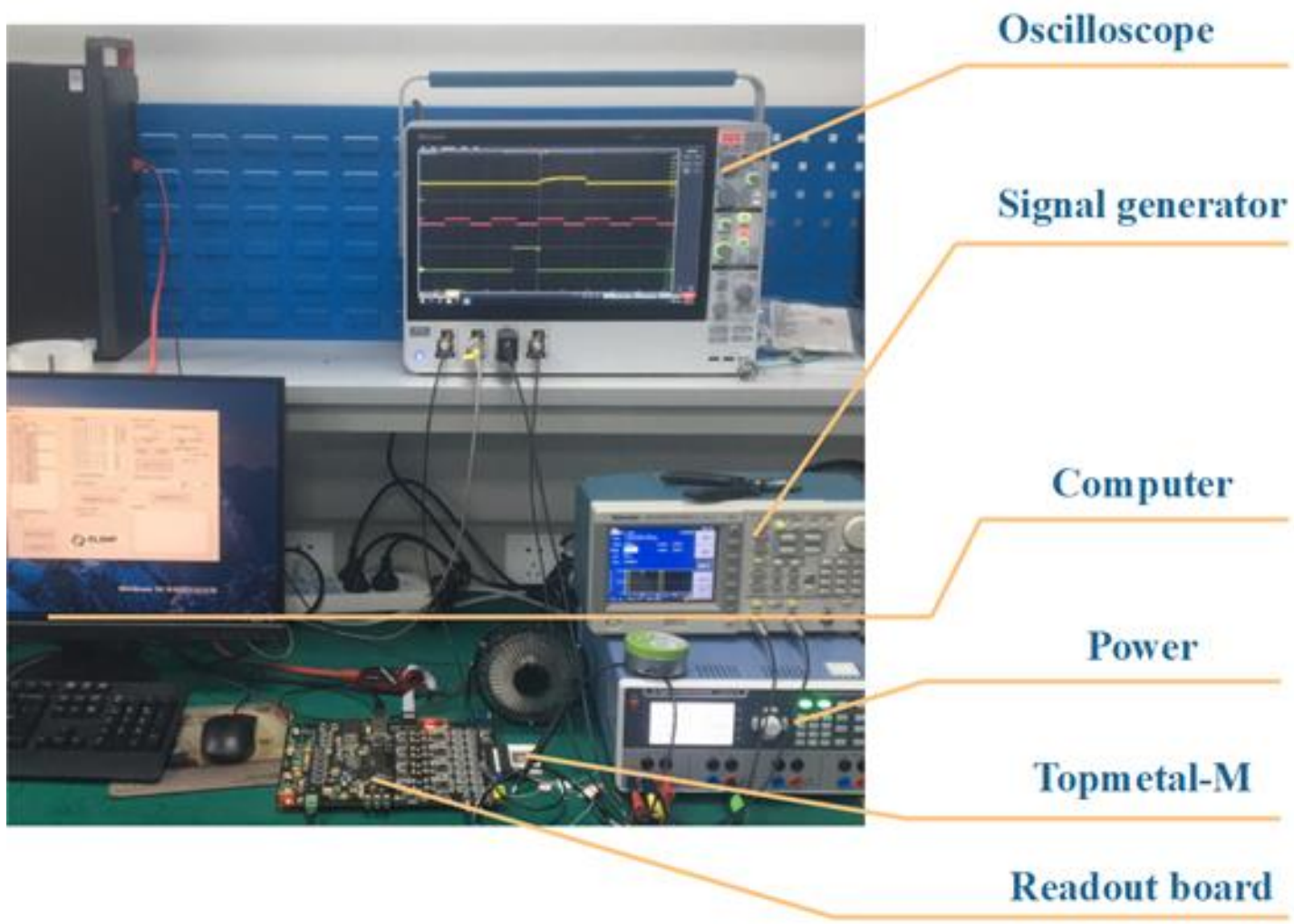

Fig. 13 The Topmetal-M test platform.

### 4.2. *Analog noise measurement*

The scan module was configured to work at the mode of stopping at one fixed pixel. All the following measurement results were obtained from this pixel. The CSA output waveform was sent to the analog readout channel and sampled by the 14-bit ADC on the readout board. An electrical pulse was applied on the guard-ring. Since there is a coupling capacitance ($C_{couple}$) between the guard-ring and the *Topmetal* sensor, the electrical pulse is injected into the $C_{couple}$ to provide a charge signal to the CSA.

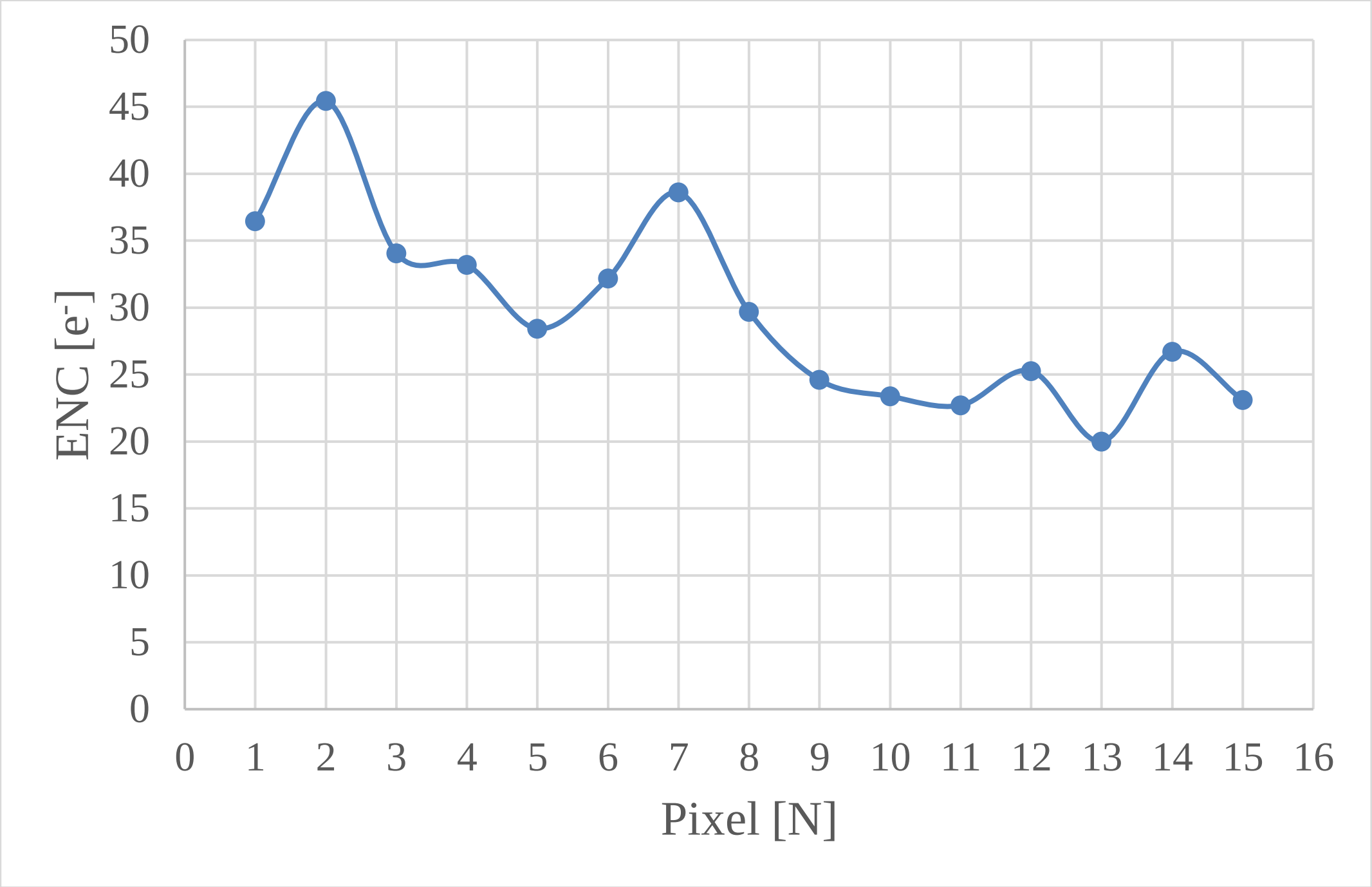

Fig. 14 ENC of the tested pixels.

The noise performance is measured by recording many analog readout output pulses with a fixed injected voltage ΔV through the PAD guard-ring. In our tests, the ΔV is 200 mV and the input charge can be calculated as Qin = $C_{couple}$ • ΔV / e = 5875 $e^-$. In total, 15 pixels have been randomly selected in the tests. For each pixel, the output of the CSA has been recorded for 1000 times. μ (the mean value of the pulse height) and σ (the fluctuation of the pulse height) are extracted from the results. The ENC is calculated as $Q_{in}$ • σ / μ. The ENC of the 15 tested pixels are shown in Fig. 14, which is in the level of 20 $e^-$ and 50 $e^-$. This is in accordance with the simulation results.

*4.3. TAC function test*

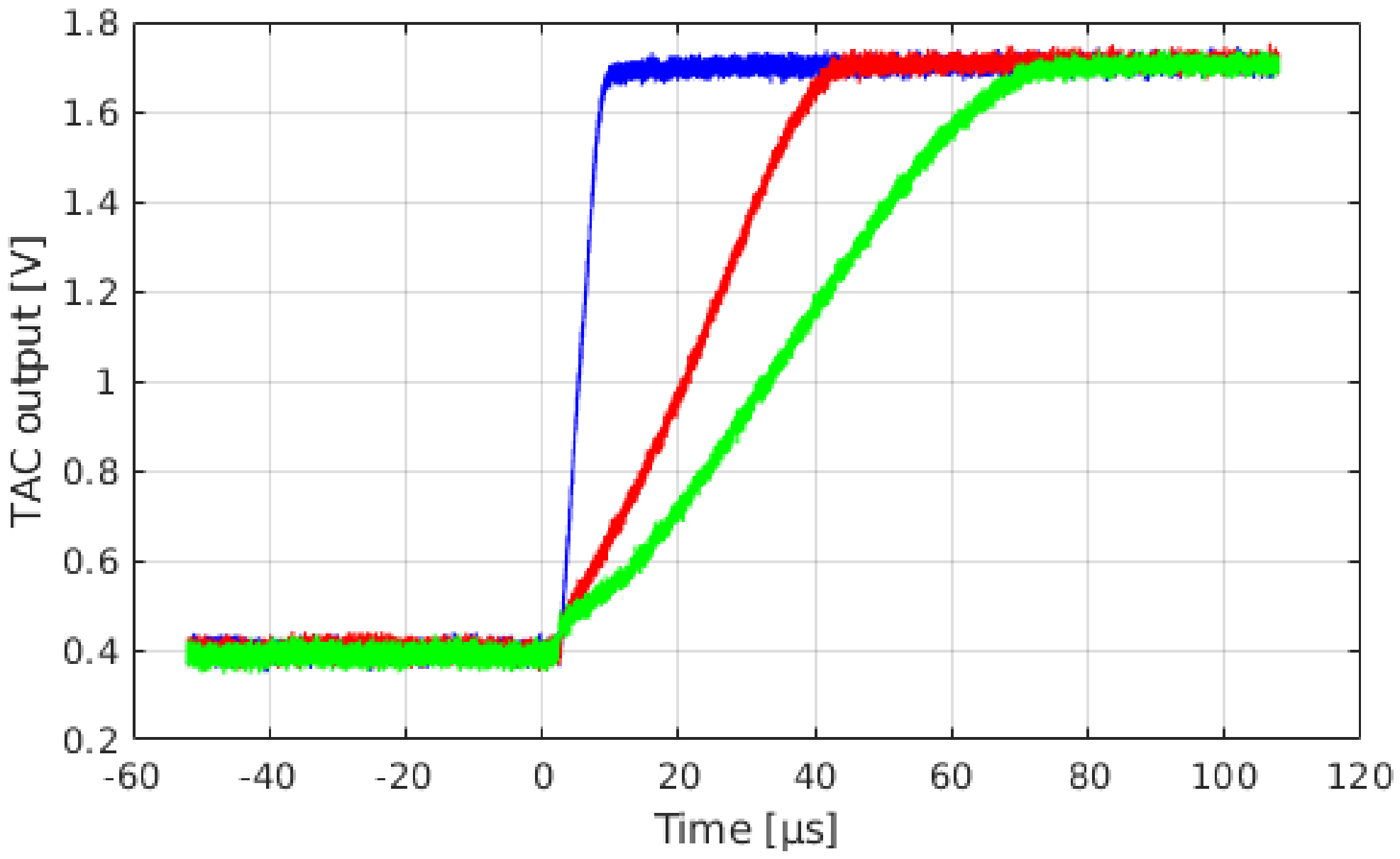

Fig. 15 The signal waveforms in the TAC test.

While testing the TAC, the bias voltage of the current source to charge the TAC was set to be 2.648 V, 2.748V and 2.798 V, respectively. Fig. 15 shows the waveforms of the TAC output, where the output amplitude is 1.314V. The dynamic range are 9.565 µs, 46.224 µs and 75.286 µs, respectively. The time resolution is calculated to be 0.89 ns, 4.29 ns and 6.99 ns, respectively. The measurement range and timing resolution obtained are in accordance with the simulation in section 3.2, but the bias voltage of the current source is about 0.15 V lower than is in the simulation.

## 5. Conclusion

The Topmetal-M sensor is a novel pixel sensor, which combines the Monolithic Active Pixel Sensor and the *Topmetal* sensor. It collects the ionized charge deposited by the particle in the sensor and along the track over the sensor. Hence, it can reconstruct the trajectory of the incident particle and measure the position of the particle hit on the sensor. With the in-pixel circuits, the Topmetal-M can record the energy and arrival time of the particles. The Topmetal-M has 400 × 512 square pixels, each single of which is in the size of 40 µm × 40 µm. The Topmetal-M is fabricated in a 130 nm high-resistivity CMOS technology, and aims to verify the feasibility of this process for pixel sensors. Lab tests performed on the Topmetal-M with reverse bias are reported. The ENC of the 15 randomly selected pixels in the energy measurement is between 20 $e^-$ and 50 $e^-$, which is in accordance with the simulation. With the TAC bias voltage set to be 2.648 V, 2.748V and 2.798 V, dynamic range of the timing measurement is tuned to be 9.565 μs, 46.224 μs and 75.286 μs, respectively. This corresponds to the timing resolution of 0.89 ns, 4.29 ns and 6.99 ns, respectively. Further test will be performed on this Topmetal-M to guide the design of the next version, in which the pixel pitch will be scaled-down and high-resolution column-parallel ADCs will be included.

## Acknowledgment

The research is supported by the Strategic Priority Research Program of Chinese Academy of Sciences, Grant No. XDB34000000, the National Science Foundation of China under Grant Nos. 11605071, 11705065, 11805080, 11975292, 11927901 and 11875304, the National Key Research and Development Program of China under Grant Nos. 2016YFA0400404 and 2016YFE0100900 and the CAS Pioneer Hundred Talent Program.